\documentstyle[12pt]{article}
\begin{document}
\begin{center}
\Large
Coherent solitary-wave of mixing layer turbulence in the physical-plus-eddy space  \\
\vspace{3mm}
%%%Shunichi Tsug\'{e}\footnote{E-mail: mv118@mel.go.jp}\\
Shunichi Tsug\'{e}\footnote{phone \& fax: 81-298-47-6667; email: shunt@tara.tsukuba.ac.jp}\\
\end{center}
\begin{center}
Tsukuba Advanced Research Alliance, University of Tsukuba, Ibaraki 305 Japan\\
{\Large and\\}
\end{center}
\begin{center}
\Large
Burtsitsig Bai\footnote{phone: 81-298-58-7083; fax: 81-298-58-7091; email: mv118@mel.go.jp}\\
\end{center}
\begin{center}
Fluid Engineering Division, Energy Engineering Department, Mechanical Engineering Laboratory, 
1-2 Namiki, Tsukuba, Ibaraki 305 Japan
\end{center}
\setlength{\baselineskip}{6mm}
\noindent {\bf Abstract}\\

A six-dimensional Navier-Stokes equation derived by one of the authors(ST) is solved for a turbulent mixing layer to demonstrate that it has a solitary wave solution. Turbulence intensities and Reynolds' stress are calculated using this solution, showing satisfactory agreement with experiments although no emperical constants are involved in the theory.\\
Key words: turbulence, solitary wave, mixing shear layers\\

With their recent advent parallel/superparallel computers have provided a powerful tool for computational fluid dynamicists. With regards to direct numerical simulation(DNS) of turbulence, however, the gap between the Reynolds number $10^4$ that can be reached using those computers is still far short of the target ones $10^7\sim10^8$ to be practical with airplane designers. This computational difficulty reflects the fact that with increase in the Reynolds number $R$, grid size required for DNS becomes finer as $R^{-3/4}$. This situation is caused by the fractal structure of turbulence\cite{Sreen}\cite{SreAnu} that the Kolmogorov scale($\sim R^{-3/4}$) is the lower fractal limit above which turbulent fluid quantities are not considered as smooth, but are self-similarly irregular with fractal dimension of about 2.36. This is where differentials are not to be replaced with finite differences, or desperately large number of Fourier modes need to be considered in employing the spectral method. 
To remedy this `small eddy difficulty' of high Reynolds number turbulence, it is advisable to work with averaged fluid variables rather than fractal ones directly. Of two candidate methods to meet this purpose, namely, the large-eddy simulation and the statistical theory of turbulence, the latter which is of classical origin\cite{Karman} is reexamined on the basis of first principles of non-equilibrium statistical mechanics\cite{Tsuge1} $\sim$ \cite{Tsuge5}: It has led us to an averaged Navier-Stokes equation in a six-dimensional(6D) space comprising 3D physical plus 3D eddy spaces\cite{Tsuge6}.

This formalism allows us to deal with smoothed quantities at the expense of increased number in independent variables. Furthermore it has a computationally favorable feature that the small eddies are mapped onto a small region around the origin of the eddy coordinates. Thus the small eddy difficulty is eliminated by employing fine mesh only in the vicinity of the origin of the eddy space, not necessarily the whole computational domain as in the DNS.

The objective of this paper is to check if the equation thus derived has a solution that is physically sound by solving a problem of practical interest. A turbulent mixing shear layer serves as such an example.

Since the discovery of the coherent structure of turbulence\cite{Brown}, this problem has attracted attentions of computational fluid dynamicists as well as experimenters. DNS, however, has not been able to reach Reynolds numbers high enough to establish spatial self-similarity. Efforts are then directed to solving a similar flows of tangentially opposing parallel streams where turbulent mixing layer grows timewise, not streamwise. Reliable DNS result is reported up to $R=2\times10^4$\cite{Rogers}. This flow configuration is much easier to compute(average flow does not change streamwise). This parallel mixing layer, however, is fictitious in nature(no experiments possible), and the similarity to the spatially growing mixing layer is only qualitative(no coordinate transformation of equations and boundary conditions from one to the other is possible). Therefore for quantitative comparison with existing data to be possible, we need to solve spatially growing case directly. 

Experimentally a turbulent mixing layer is generated between two parallel streams with different velocities $U_{\pm\infty}$, starting to merge at $x_1=x_2=0$. It is shown \cite{Oster}\cite{Bell} that 2-D incompressible mixing layers tend to be self-similar with increase in thickness $\delta$ of the mixing region that grows linearly with streamwise distance; $\delta=\alpha x_1$. This experimental evidence indicates that the governing equations [Eqs.(34) and (35) of ref.9]
\begin{eqnarray}
\partial_j Q_j=0
\end{eqnarray}
\begin{eqnarray}
(\frac{\partial}{\partial t}-C_\ell\frac{\partial}{\partial S_\ell}+u_\ell\partial_\ell-\nu\partial_\ell^2)Q_j+\frac{1}{\rho}\partial_jQ_4+\frac{\partial u_j}{\partial x_\ell}Q_\ell+\partial_\ell Q_jQ_\ell=0 
\end{eqnarray}
where $Q$ replaces $\rho^{-1}q$($\rho$ ; the density) of ref.9, and
\begin{eqnarray}
\partial_j\equiv\frac{\partial}{\partial x_j}+\frac{\partial}{\partial S_j}
\end{eqnarray}
are controlled by reduced number of self-similar variables
\begin{eqnarray}
\eta=x_2/\delta,\;\;\;
\mbox{\boldmath$s$}=\mbox{\boldmath$S$}/\delta
\end{eqnarray}
where subscripts 2 and 3 refer to normal and spanwise directions viewed from the splitter plate separating the two parallel streams. This observation of self-similarity together with the group-theoretical consideration \cite{Birkhoff} tell us that terms with viscosity be negligible, corresponding to the physical situation that the viscous stress is overwhelmed by the Reynolds stress everywhere in the flow. 

Even under this simplifying condition, the equations have four independent variables to which existing computational methods are not applicable directly. To make them tractable it is proposed to suppress one of the eddy variables; $\partial/\partial s_2=0$. This choice saves the minimum requisite that longitudinal vortices depend on $s_3$ and transverse ones on $s_1$ more strongly than others. Under these conditions  Eqs.(1) and (2) reduce to the following set 
of `inviscid' equations:
\begin{eqnarray}
\partial_1 q_1 + \frac{\partial q_2}{\partial \eta} + \frac{\partial q_3}{\partial s_3}=0
\end{eqnarray}
\begin{eqnarray}
\underline{NL}\; q_1+\partial_1 q_4-\alpha \eta u'q_1+u'q_2=0
\end{eqnarray}
\begin{eqnarray}
\underline{NL}\; q_2+\frac{\partial}{\partial\eta}q_4-\alpha \eta v'q_1+v'q_2=0
\end{eqnarray}
\begin{eqnarray}
\underline{NL}\; q_3+\frac{\partial q_4}{\partial s_3}=0
\end{eqnarray}
with
\begin{eqnarray}
\left.
\begin{array}{ll}
\partial_1 \equiv (1-\alpha s_1)\displaystyle\frac{\partial }{\partial s_1}-\alpha ( \eta \frac{\partial }{\partial \eta}+s_3\frac{\partial}{\partial  s_3})\\
\underline{NL} \equiv   \displaystyle\frac{\partial }{\partial t}-c \frac{\partial }{\partial s_1}+(u+q_1)\partial_1 + (v+q_2)\frac{\partial}{\partial\eta}+q_3\frac{\partial}{\partial s_3} 
\end{array}
\right\}
\end{eqnarray}
where $(u(\eta),v(\eta),0)$ is the mean velocity vector made nondimensional using the velocity difference $U=U_{\infty}-U_{-\infty}$ ,and $q_\alpha$ and $c$ are nondimensional fluctuations and phase velocity defined similarly;
\begin{eqnarray}
\left.
\begin{array}{ll}
q_j = Q_j/U \;\;\;\;\; q_4 = Q_4/ {\rho U^2}   \\
c = C/U\:.
\end{array}
\right\}
\end{eqnarray}

Since this paper is focused on a quick check of the adequacy of the proposed approach, in particular, the existence of solitary wave solution, we do not solve, but instead adopt measured values for $u$, while $v$ is computed using the continuity equation;
\begin{eqnarray}
v= \int_0^{\eta} \eta \frac{du}{d \eta} d\eta .
\end{eqnarray}

The phase velocity  $\bf\it\mbox{\boldmath$c$}$, namely, the propagation velocity of turbulent vortices in the $(\mbox{\boldmath$x$},\mbox{\boldmath$s$})$ space is yet to be specified. Whether it is to be given on first principles is not well understood at this stage. Therefore we will explore this point by invoking two different assumptions representing two extreme cases of large and small eddies, and will see how they compare to each other.

The one is to put 
\begin{eqnarray}
c = \frac{U_\infty +U_{-\infty }}{2U}(\equiv u_0)
\end{eqnarray}
whose right-hand-side represents the phase velocity of vortices as observed by visualisation experiments, in other words, the propagation velocity of large eddies\cite{Bernal}. The other is to put 
\begin{eqnarray}
\mbox{\boldmath $c$}=\mbox{\boldmath $u$}  \label{eqn:c_tay}
\end{eqnarray}
which represents Taylor's hypothesis that claims the vortices to be carried with local mean flow. It is more likely that small eddies obey (13) rather than (12).

Initial-value problems are solved employing the pseudo-compressibility method\cite{Bai} in the 3-D space $(\eta,s_1,s_3)$. The method assumes a fictitious term \( \beta^{-1} \partial f_4/{\partial t} \;\;\; (\beta \gg 1) \) 
added on the continuity equation (5) to allow the whole set of equations (5) through (8) for time evolution procedure.

A unique solitary wave solution is shown to build up eventually independent of whatever initial distributions to start with. The solution having the property
\begin{eqnarray}
q_1,q_2,q_4 & ; & \mbox{ even in }s_3,\nonumber\\
q_3 & ; & \mbox{ odd in }s_3\nonumber
\end{eqnarray}
is sought to meet with the two-dimensional structure of the mean flow; $\langle u'_1u'_3 \rangle=\langle u'_2u'_3 \rangle=0$. It is also confirmed that the 
time-dependent method leads asymptotically to a steady-state solution, in other words, that there are no signs for the solution to have a limit-cycle or strange-attractors.

The mean flow condition employed for the computation is 
\begin{eqnarray}
U_{-\infty}/U_\infty = 0.6
\end{eqnarray}
the same as the experimental condition of refs.\cite{Oster} and \cite{Bell}. The mean velocity profile for $u$ is taken from ref.\cite{Oster}, and that for $v$ is calculated from Eq.(11).

In Figs.1 are shown the birds-eye-view representation of the solitary wave solution for $q_1(\eta,s_1,s_3)$ projected onto the plane of $\eta = 0$. The computation is based on two different assumptions(12)(Figs.1a) and(13)(Figs.1b) for the phase velocity. They look quite similar at least qualitatively. In both streamwise($s_1$) and spanwise($s_3$) directions of eddy coordinates, the main peak is accompanied by valleys, which is the sign of existence of adjacent counter-rotating vortices.

Once the solitary-wave solution has been obtained the fluctuation correlations and the turbulence intensities including the Reynolds stress are calculated through the following integrals 
\begin{eqnarray}
\langle u'_j(\mbox{\boldmath$x$})u_\ell(\widehat{\mbox{\boldmath$x$}})' 
\rangle & \equiv & R_{jl}(\mbox{\boldmath$x$},\widehat{\mbox{\boldmath$x$}})\nonumber \\
& = & \frac{1}{(2\pi)^2}\int\int^\infty_{-\infty} 
ds_1ds_3 q_j(\eta,s_1,s_3)q_\ell(\widehat{\eta}, 
s_1+\frac{\widehat{x}_1-x_1}{\delta}, 
s_3+\frac{\widehat{x}_3-x_3}{\delta})\;\;\;\;
\end{eqnarray}
which is the 2-D and self-similar version of formula(39) of ref.9. Figs.2 show the turbulence instensities obtained by putting $\widehat{\mbox{\boldmath$x$}}=\mbox{\boldmath$x$}$ in (15): Those figures correspond to a) streamwise $(j=\ell=1)$, b) transverse $(j=\ell=2)$, c) spanwise $(j=\ell=3)$ intensities of velocity fluctuations and d) Reynolds stress $(j=1,\ell=2)$. Two curves with solid lines correspond to respective assumptions for the phase velocities as indicated in the figures. It is observed that both theoretical curves fit better with Bell-Mehta's data rather than Wygnanski's. For each case they are reasonably close to each other considering that the two assumptions stand for the opposite extreme of
the eddy size. This computational evidence implies robust tendency of the governing equations towards the solitary wave formation. As is easily confirmed, the governing equations do not allow for solutions that are symmetric with respect to $\eta$. It may well be expected that the seemingly symmetrical distribution as observed in experiments makes its appearance only after the integration over $\mbox{\boldmath$s$}$-space has been effected (formula (15)). The sensible asymmetry recognized in computed results may be attributable to insufficient mesh numbers as limited by the memory size of the computer. 

Fig.3 shows the vertical turbulent transport of the Reynolds stress $R_{122}(\eta) = <u_1' u_2' \widehat{u_2}'>$ for $c=u_0$ and compared with experiment\cite{Bell}. The agreement between theory and experiment for such a small and subtle quantity is unexpectedly close to each other.

Fluctuation-correlations of the vertical velocity component $R_{22}=\langle u'_2\widehat{u}'_2\rangle$ using (15) in streamwise (a;$\eta=0,\widehat{x}_3=x_3$) and spanwise (b;$\eta=0,\widehat{x}_1=x_1$) directions, respectively, are computed and plotted in Figs.4. Each of these figures corresponding to the two postulates, $c=u_0$ and $c=u$, falls very close to each other, so only the case $c=u$ is shown here. Both correlation curves have the points of zero-crossings, indicative of adjacent vortices of opposite rotations, also of the clearcut existence of bulk of co-rotating cores in both directions. Length $\ell_3$ of zero-crossing in the spanwise correlation curve is available from Bell-Mehta's experiment\cite{Bell}, falling quite close to the predicted one.

\clearpage
\begin{center}
\Large
Figure Captions\\
\end{center}

\vspace{10mm}
Figure 1: Solitary wave solution of $q_1(\eta,s_1,s_3)$ on two different assumptions for phase velocity $c$ [(a) and (b)] viewed on the plane of $\eta = 0$.\\

\vspace{10mm}
Figure 2: Distributions of turbulent intensities in transverse ($\eta$ -)direction, compared with data by Oster-Wygnanski[12] and Bell-Mehta[13]; a)streamwise ($R_{11}$), b)transverse ($R_{22}$), c)spanwise ($R_{33}$) and d)Reynolds' shearing stress ($R_{12}$).\\

\vspace{10mm}
Figure 3: Transverse transport of Reynolds' stress ($R_{122}$) and comparison with data by Bell-Mehta[13].\\

\vspace{10mm}
Figure 4: Stream- and span-wise variation of correlation of the vertical velocity component $R_{22}$ at $\eta$ = 0.\\

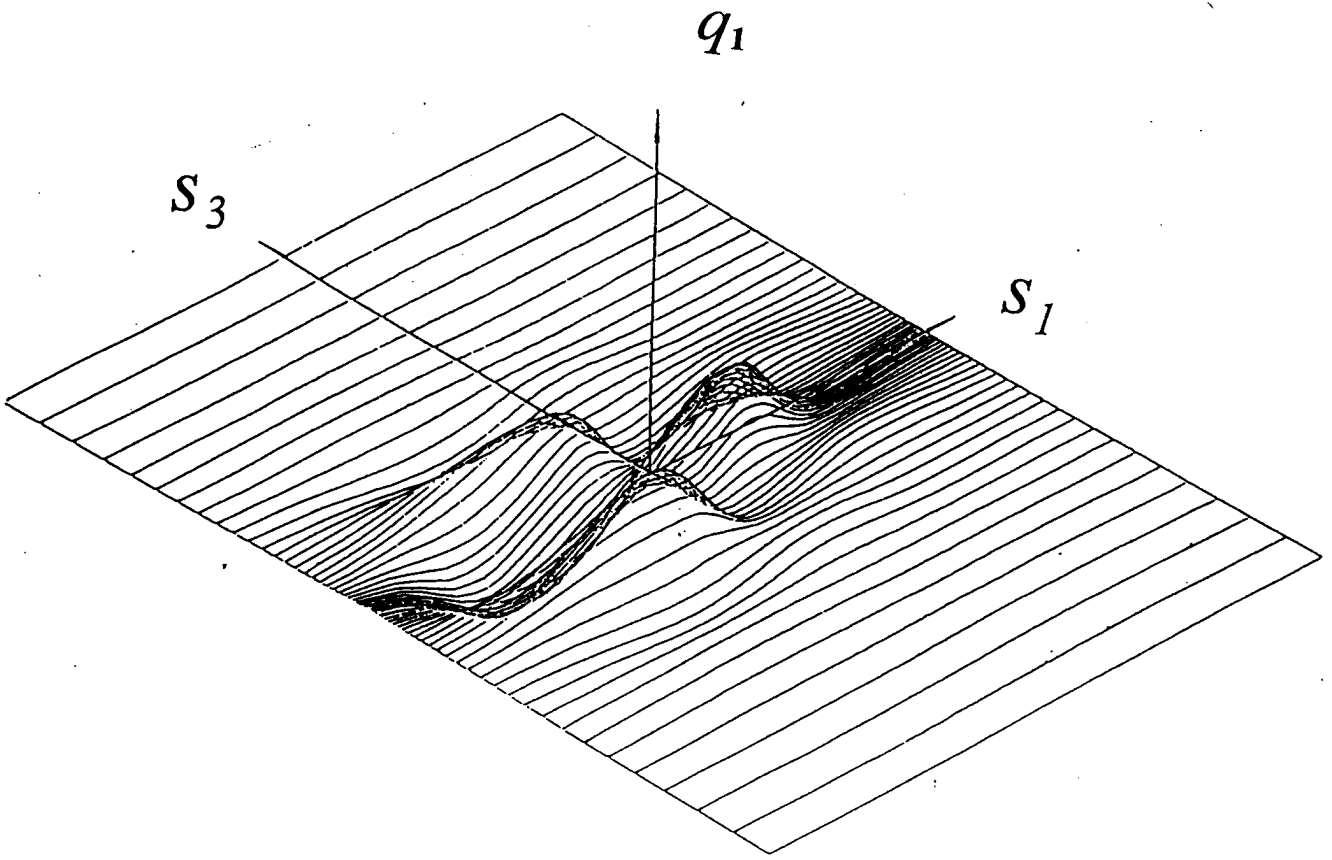
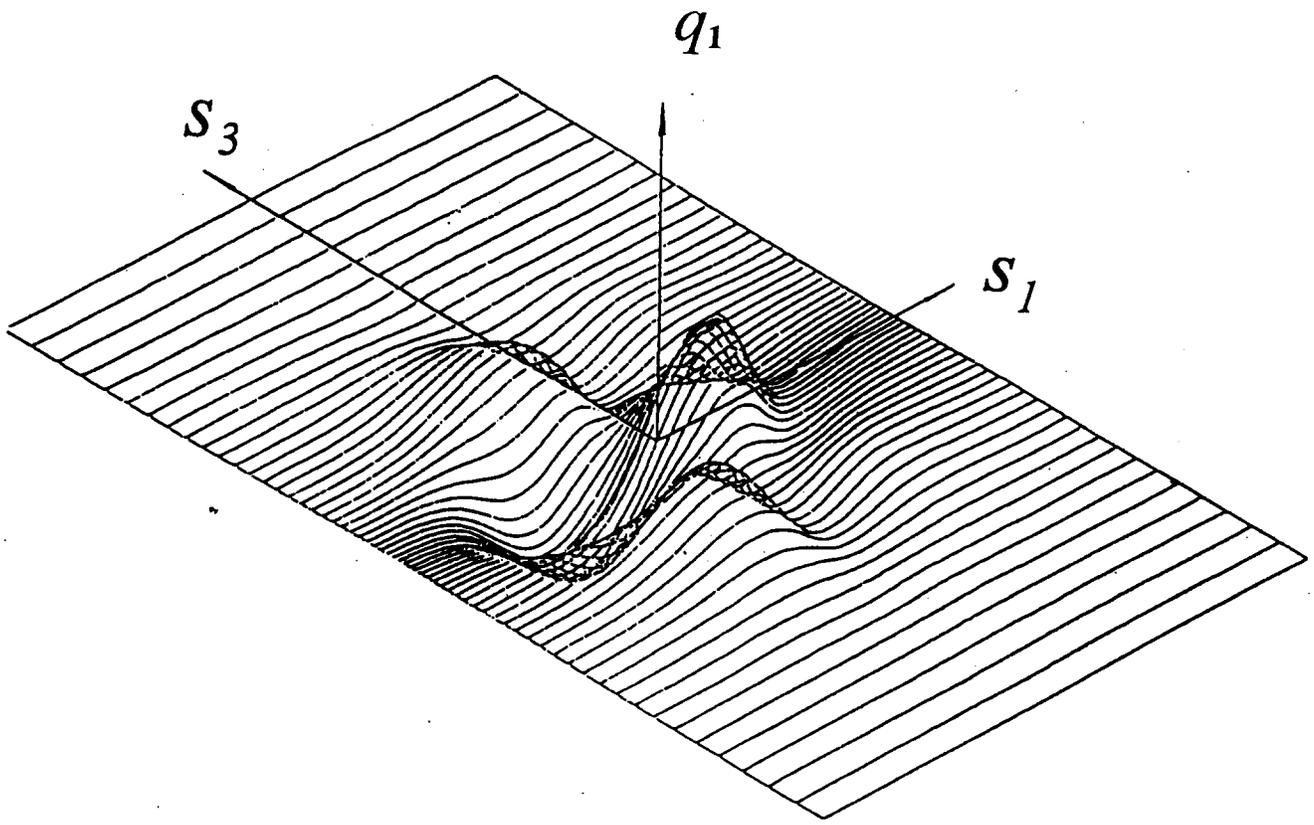
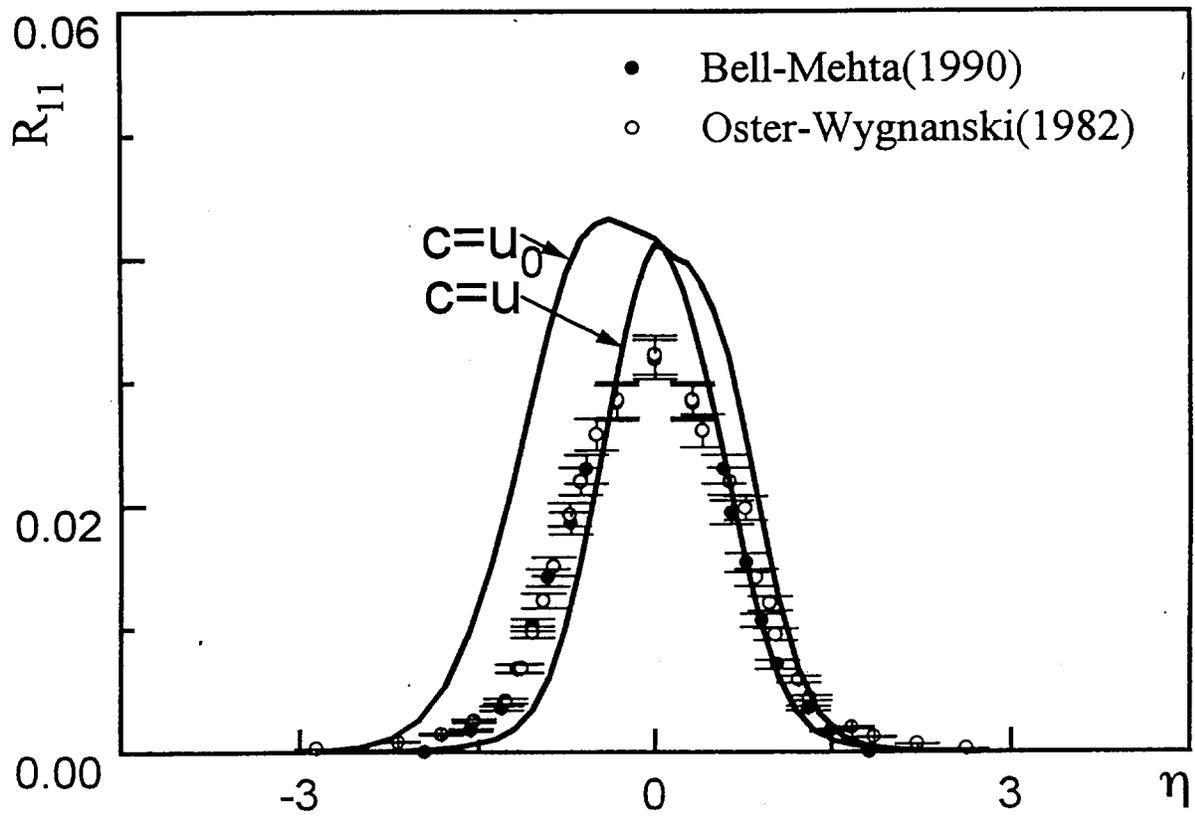
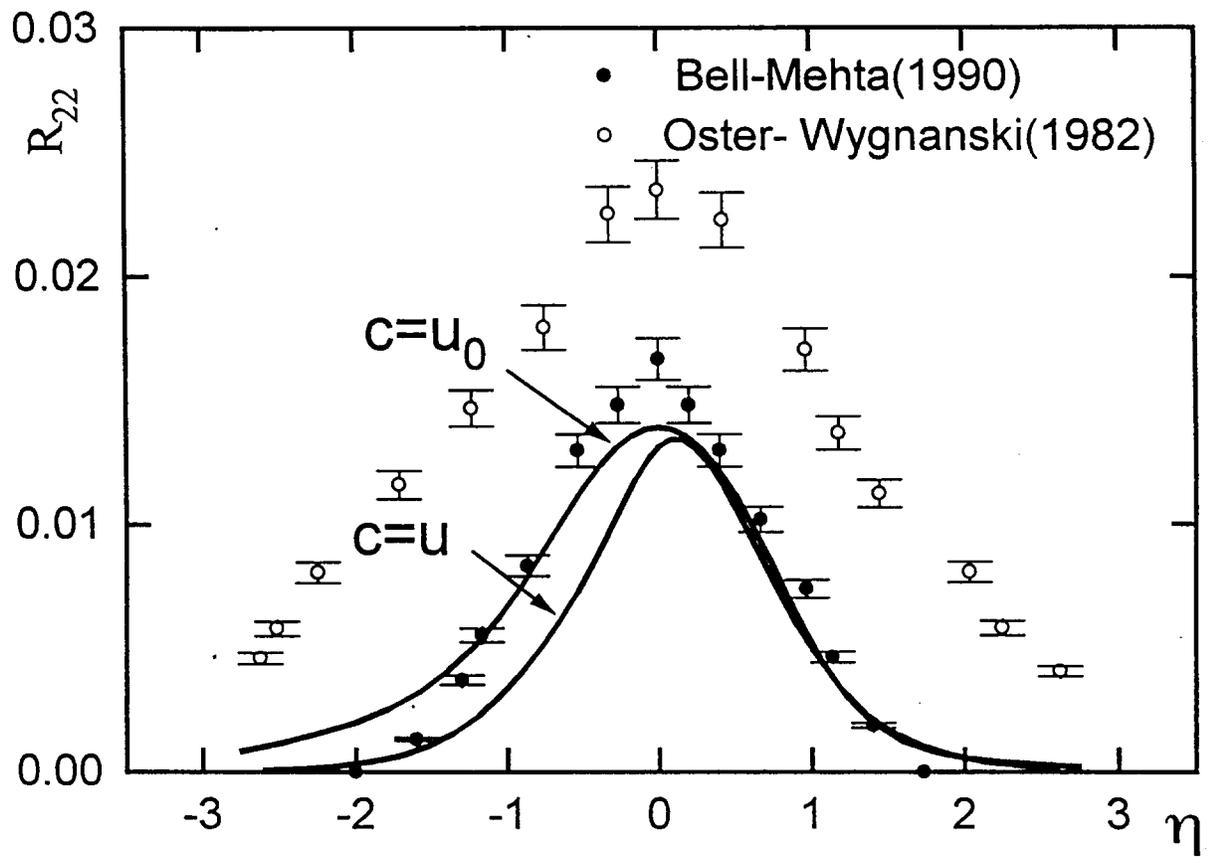
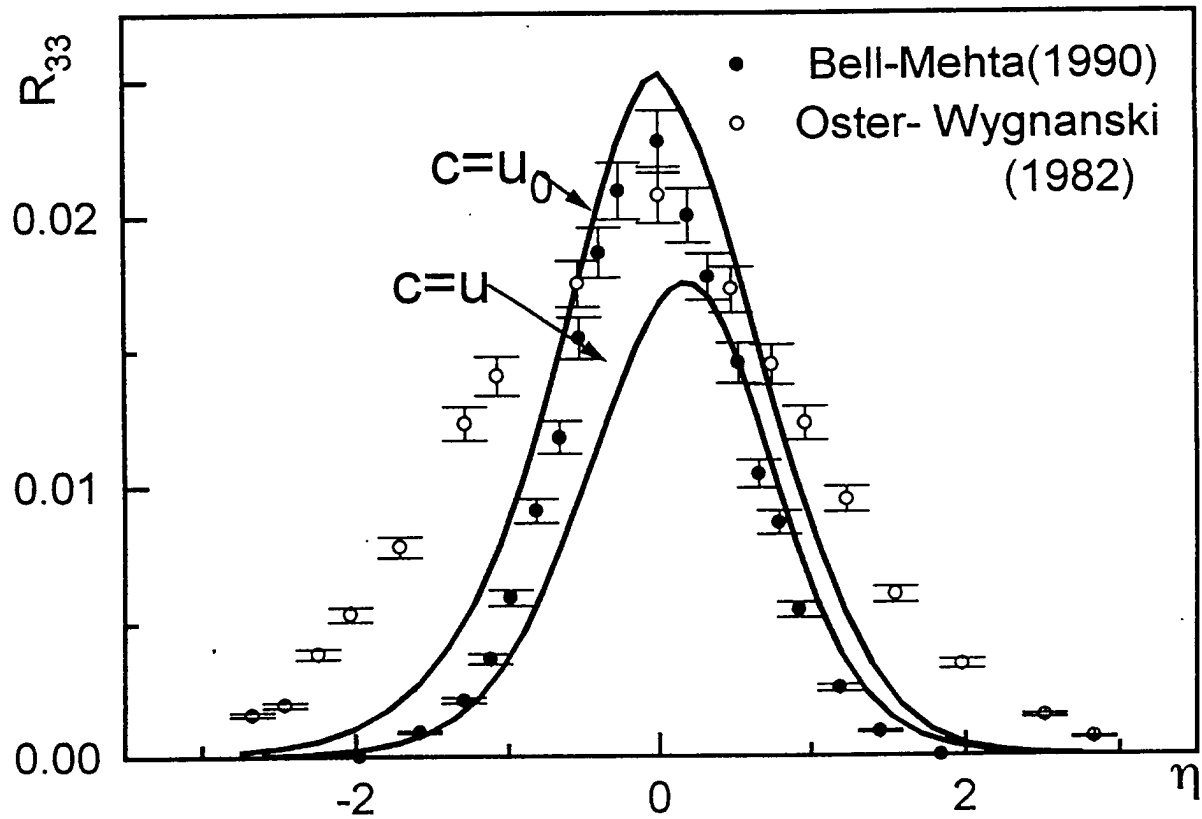
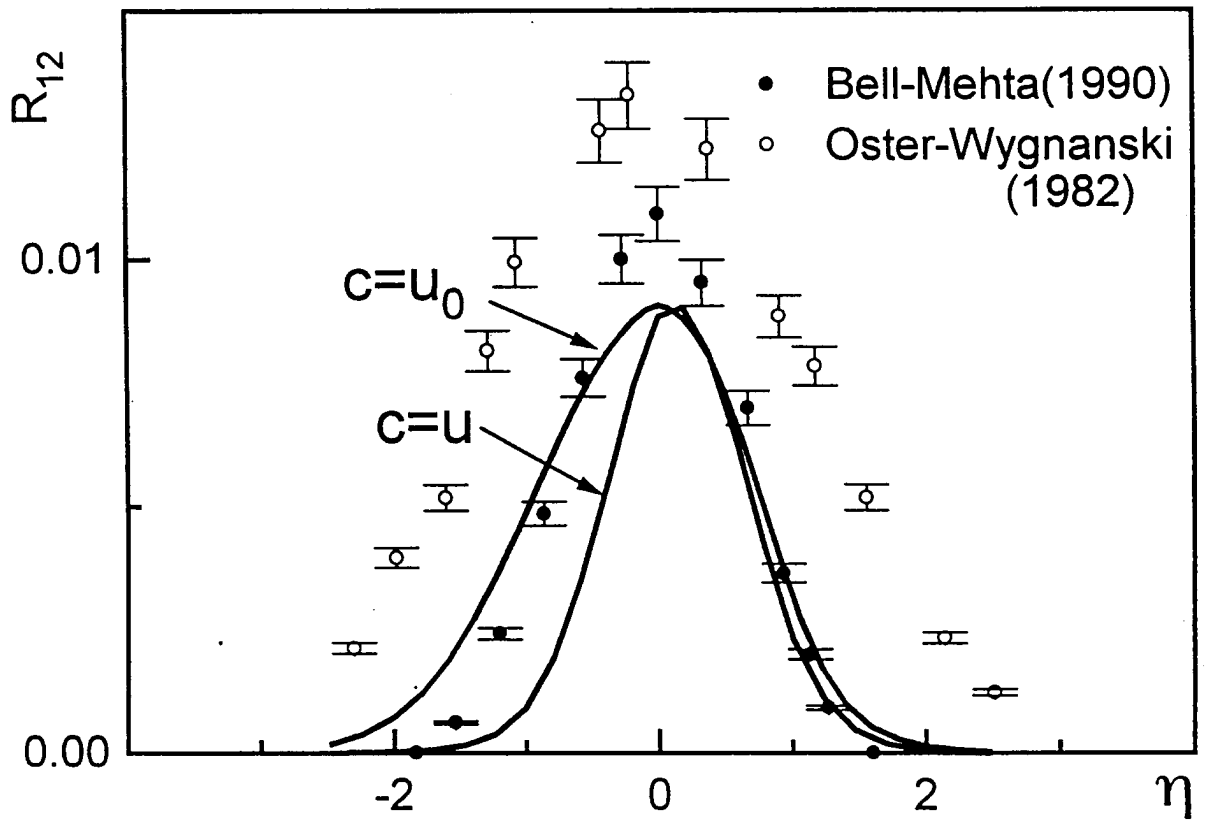
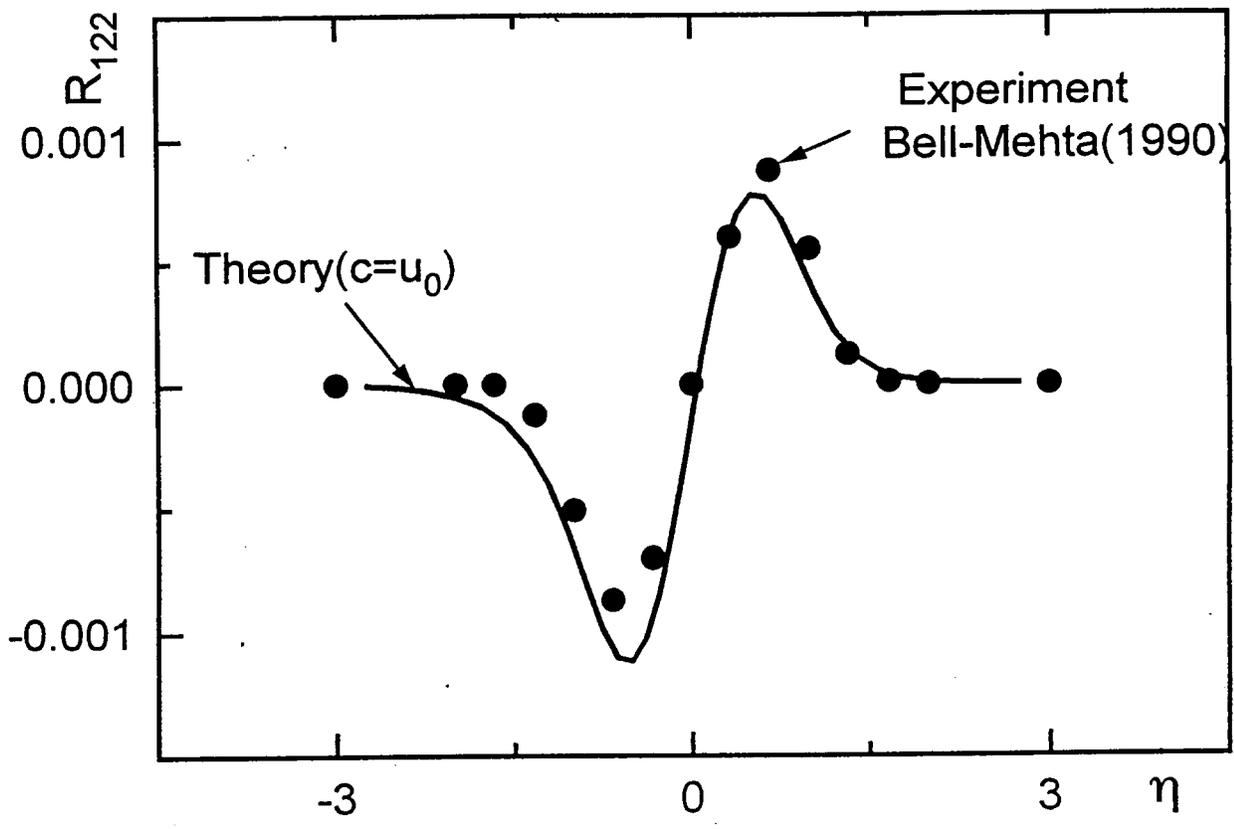
\end{document}